\documentclass[slac_one]{revtex4}
\usepackage{graphicx}
\usepackage{fancyhdr}
\newcommand{\enull}{\mbox{$E_0$}}

\newcommand {\be}{\begin{equation}}
\newcommand {\ee}{\end{equation}}
\newcommand {\bea}{\begin{eqnarray}}
\newcommand {\eea}{\end{eqnarray}}
\newcommand{\bdec}{\mbox{$\beta$ decay}}

\newcommand{\belecs}{\mbox{$\beta$-electrons}}

\newcommand{\mtwo}{\mbox{$m_\nu^2$}}
\newcommand{\dmtwo}{\mbox{$\Delta m_\nu^2$}}

\newcommand{\NnCL}{\mbox{95\%\,C.L.}}

\newcommand{\etal}{\mbox{\it et al.},}

\newcommand{\AmS}{{\protect\the\textfont2
  A\kern-.1667em\lower.5ex\hbox{M}\kern-.125emS}}

\begin{document}

\title{The KATRIN Neutrino Mass Experiment} 

%

\author{J. Wolf (for the KATRIN collaboration)}
\affiliation{Institut f\"ur experimentelle Kernphysik,
Universit\"at Karlsruhe, Postfach 6989, 76128 Karlsruhe, Germany}

\begin{abstract}

The {\bf Ka}rlsruhe {\bf Tri}tium {\bf N}eutrino experiment
(KATRIN) aims to measure the mass of electron neutrinos from
beta-decay of tritium with an unprecedented sensitivity of $0.2\,
eV/c^2$ improving present limits by one order of magnitude. The
decay electrons will originate from a $10\, m$ long windowless,
gaseous tritium source. Super-conducting magnets guide the
electrons through differential and cryogenic pumping sections to
the electro-static tandem spectrometer (MAC-E-filter), where the
kinetic energy will be measured. The experiment is presently being
built at the {\em Forschungszentrum Karlsruhe} by an international
collaboration of more than 120 scientists. The largest component,
the $1240\, m^3$ main spectrometer, was delivered end of 2006 and
first commissioning tests have been performed. This presentation
describes the goals and technological challenges of the experiment
and reports on the progress in commissioning first major
components. The start of first measurements is expected in 2012.

\end{abstract}


\maketitle

\thispagestyle{fancy}

\section{Introduction}

The Standard Model of particle physics describes neutrinos as
massless, neutral fermions, which can be detected only via weak
interactions. The observations of neutrino flavour oscillation
\cite{Osc04} indicate the existence of massive neutrinos. Since
experiments investigating these neutrino oscillations are only
sensitive to differences of the square of mass eigenvalues
$\dmtwo=|m^2(\nu_i)-m^2(\nu_j)|$ and not to the absolute values,
the scale and hierarchy of neutrino masses are not determined yet.
An identification of the mass scheme realized in nature would not
only solve the puzzle of absolute neutrino masses but could also
point to the mechanisms of mass generation in extensions of the
Standard Model.

In standard cosmological models, our universe is filled with
primordial neutrinos arising from freeze-out in the early
universe. These neutrinos are natural candidates for non-baryonic
hot dark matter. Depending on the actual neutrino mass, the
neutrino content of the universe can exceed the baryonic mass
density.

The above arguments demonstrate the importance of the absolute
neutrino mass scale for both particle physics and cosmology. A
model-independent approach to determine the neutrino mass is the
kinematical analysis of electrons from radioactive \bdec\ near the
endpoint energy \enull . A non-vanishing neutrino mass reduces the
electron endpoint energy and distorts the shape of the electron
spectrum in the vicinity of $\enull - m(\nu)$.

\section{KATRIN Experiment}

\begin{figure}[ht]
 \centerline{\includegraphics[width=12cm]{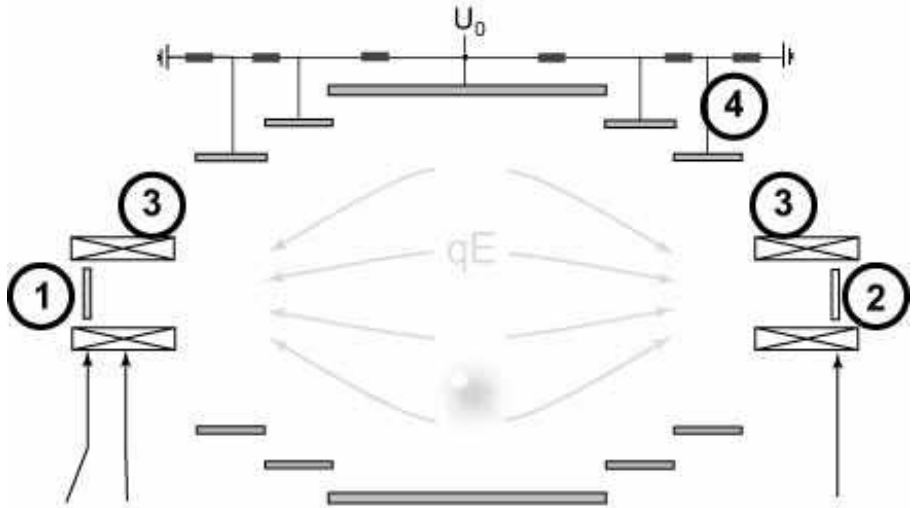}}
 \caption{Scheme of a MAC-E-filter with tritium source (1), the electron detector (2), the
 superconducting solenoids (3) and the HV electrode system (4).}
 \label{fig-MAC-E-filter}
\end{figure}

The scheme of a measurement investigating the endpoint region of
the electron energy spectrum from tritium \bdec\ ($^3{\rm H}
\rightarrow\ ^3{\rm He} + e^- + \bar{\nu}_e$) with an
electrostatic filter is illustrated in
figure~\ref{fig-MAC-E-filter}.

Electrons emitted by \bdec\ in the source are guided magnetically
through the electrostatic retarding potential of the spectrometer,
to be counted by a segmented silicon detector. This {\em magnetic
adiabatic collimation} followed by the {\em electrostatic} (MAC-E)
filter represents an integrating high pass filter for \belecs.
This technique has been used in the Mainz and Troitsk experiments
with different tritium sources, providing the most stringent,
model-independent limits on the neutrino mass so far, with
(\cite{Mainz},\cite{Troitsk},\cite{Kra04},\cite{Lob03})
\begin{equation}
m(\nu_e)<2.2 \, eV/c^2 \,(\NnCL\footnote{Confidence Limit}).
\end{equation}
With well understood systematic effects, both experiments have now
reached their sensitivity limits. The energy resolution of a
MAC-E-filter depends on the ratio of the strongest magnetic field
$B_{max}$ in the beam-line and the weakest field $B_{min}$ in the
centre of the spectrometer. Flux conservation requires that $B
\cdot A$ remains constant, where $A$ is the cross section of the
magnetic flux tube, guiding the electrons from the source to the
detector. Improving the energy resolution of such an experiment
needs a much lower $B_{min}$ and therefore a spectrometer with a
much larger cross section, which will be able to contain the
magnetic flux. An improved sensitivity requires not just size, but
also a stronger source with higher statistics and a better
understanding of systematic uncertainties.

\begin{figure}[ht]
  \centerline{\includegraphics[width=16cm]{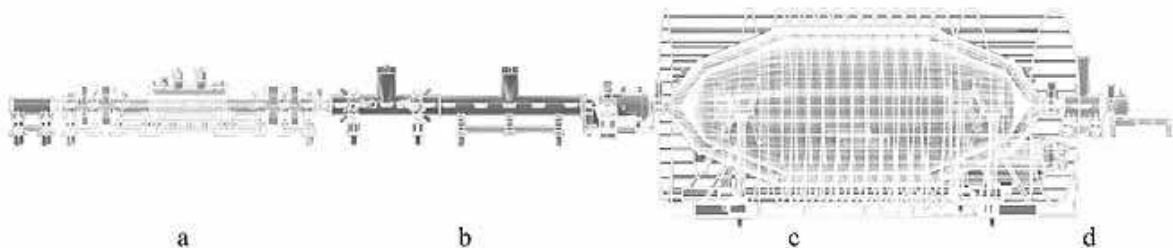}}
  \caption{\label{fig-katrin-scheme} Katrin overview: a) tritium source (WGTS),
  b) differential pumping (DPS) and cryo-pumping section (CPS),
  c) pre-spectrometer and main spectrometer, d) detector.}
\end{figure}

The {\bf KA}rlsruhe {\bf TRI}tium {\bf N}eutrino (KATRIN)
experiment will use the techniques developed in Mainz and Troitsk
with a strong gaseous molecular tritium source and an
electro-static filter of unprecedented energy resolution ($\Delta
E < 0.93\, eV$ \cite{Osi01}. The expected sensitivity for the
neutrino mass will be $0.2\,eV/c^2$ after three years of
measurement. As the experimental observable is \mtwo , an
improvement in sensitivity on $m(\nu_e)$ of one order of magnitude
corresponds to an improvement of a factor $\approx 100$ in
accuracy on \mtwo\, leading to numerous technical challenges for
this experiment. Figure~\ref{fig-katrin-scheme} outlines the
experimental set-up of KATRIN \cite{Dsgn04}, which adds up to a
total length of about $70\, m$.

The KATRIN experiment is currently under construction at
Forschungs\-zentrum Karlsruhe (FZK), Germany. With the Tritium
Laboratory Karlsruhe (TLK), a unique lab is available on site,
which will house the complete tritium inventory, including the
source and transport section. First components such as the
pre-spectrometer (2003), main spectrometer (2006), some magnets
and a prototype detector array are on site, being thoroughly
tested. The experiment is now in the construction and
commissioning phase. First measurements are expected to start in
2012, with a total measuring time of 5 years.

\subsection{Source and Transport Section}

The source and transport section includes the {\em windowless
gaseous tritium source} (WGTS), followed by {\em differential
pumping sections} (DPS) and a {\em cryogenic pumping section}
(CPS). The beam tubes inside the super-conducting solenoids have a
diameter between $75\,mm$ and $90\,mm$, interspersed with pump
ports for turbo-molecular pumps (TMP). These pumps are integrated
in the closed loop tritium circulation system. The pressure inside
the vacuum beam-line will range from $3.4 \times 10^{-3}\, mbar$
at the source to less than $10^{-11}\, mbar$ in the spectro\-meter
section.

\subsubsection{Windowless Gaseous Tritium Source}

The WGTS will be the standard $\beta$-electron source for
long-term tritium measurements. Ultra-cold molecular tritium gas
($T = 27\,K \pm 30\,mK$) will be injected through a set of
capillaries at the centre of the $10\,m$ long WGTS tube with an
injection pressure of $p_{in}=3.4 \times 10^{-3}\,mbar$ and a flow
rate of about $1.8\,mbar\,l/s$ ($40\,g(T_2)/day$). The density
profile of the gas inside the beam-tube has to be kept stable on
the $10^-3$ level. The stability depends on a constant beam-tube
temperature and inlet pressure in the WGTS. Maintaining these
conditions is a very challenging task, and the technical
feasibility will be demonstrated with the partly assembled WGTS in
2009.

After injection the T$_2$ molecules will be transported by
diffusion over a length of $5\,m$ to both ends of the source tube,
where most of the tritium will be pumped out by TMPs of the first
stage of the differential pumping section. This process leads to
an almost linear decrease of the tritium number density
\cite{Dsgn04},\cite{Sharipov04}. Electrons from $\beta$-decays
will be guided adiabatically by a magnetic field of
B$_S\,=\,3.6\,T$.

\subsubsection{Tritium Pumping System}

The background generated by tritium decay within the spectrometers
must be less than $10^{-3}$ counts/s, which limits the amount of
tritium permissible in the main spectrometer to a partial pressure
of about $10^{-20}\,mbar$. With a pumping speed of $10^6 \, l/s$
this leads to a maximum allowed tritium flow rate into the
spectrometer section in the order of $10^{-14}\,mbar\,l/s$. This
very large suppression factor will be achieved in two stages,
based on a combination of differential (DPS) and cryogenic (CPS)
pumping sections, with each stage providing a suppression factor
in the order of $10^{-7}$.

The first part of the tritium flow suppression is based on
differential pumping \cite{DPS01},\cite{DPS02},\cite{DPS03}. Along
the beam-line turbo-molecular pumps with high pumping speed reduce
the tritium flow both at the rear and front ends of the source. As
the DPS elements adjacent to the WGTS (DPS1-F, DPS1-R) influence
the stability of the gas flow in the source, they have to be
operated under almost identical conditions as the WGTS. Therefore
these first DPS sections are included in the source cryostat. The
TMPs have to be operated close to the beam-line, in order to
minimize conductance losses. Due to the strong magnets extensive
tests have been made to investigate the behaviour of TMPs in
magnetic fields.

The differential pumping system will reduce the tritium flow into
the subsequent passive cryo-trapping system (CPS) to
$10^{-7}\,mbar\,l/s$ (at 273 K). The CPS will reduce the tritium
flow by another 7 orders of magnitude, allowing only a remaining
flow below $10^{-14}\,mbar\,l/s$ into the pre-spectrometer. During
a normal measuring period of 60 days a total amount of about
$0.5\, mbar\, l$ of tritium molecules can be accumulated in the
CPS, corresponding to an activity of $1\,Ci$\footnote{The
half-life of tritium is 12.3 a}. The beam-tube of the CPS will be
kept at a temperature of $4.5\,K$. At this temperature tritium
molecules are passively adsorbed on the wall. To enhance the
trapping probability, the cold surfaces of the CPS beam-tubes will
be covered by a thin layer of argon frost. The main advantage of a
condensed gas layer, compared to a solid adsorbent like charcoal,
lies in the easy removal of both the adsorbent and the adsorbed
tritium, thus minimizing the residual tritium contamination of the
beam-tube. After each measuring period the CPS will be re-heated
and the tritium gas will be returned to the closed tritium cycle.

Although data on hydrogen and deuterium adsorption on condensed
gases are available \cite{Yuferov70}, no data on tritium
adsorption were found. The effect of electron stimulated
desorption from tritium $\beta$-decay on the migration of tritium
molecules along the cryo-sorption pump of the CPS beam-tube and
the influence on the suppression factor have been investigated in
Monte Carlo simulations \cite{CPS01},\cite{CPS05}. In order to
experimentally explore the adsorption properties of condensed
argon for tritium, the test experiment TRAP\footnote{{\bf TR}itium
{\bf A}rgon frost {\bf P}ump } has been set up by the KATRIN
collaboration and first results have been published for deuterium
and tritium \cite{CPS02},\cite{CPS03},\cite{CPS04}, confirming the
CPS design values.

All three major components of the source and transport section are
expected to be delivered between 2009 (DPS) and 2011 (WGTS and
CPS). After commissioning tests of each components they will be
connected to the spectrometer section. The source is expected to
provide first tritium electrons to be analysed in the spectrometer
early 2012.

\subsection{Tandem Spectrometer and Detector}

\begin{figure}[ht]
 \centerline{\includegraphics[width=16cm]{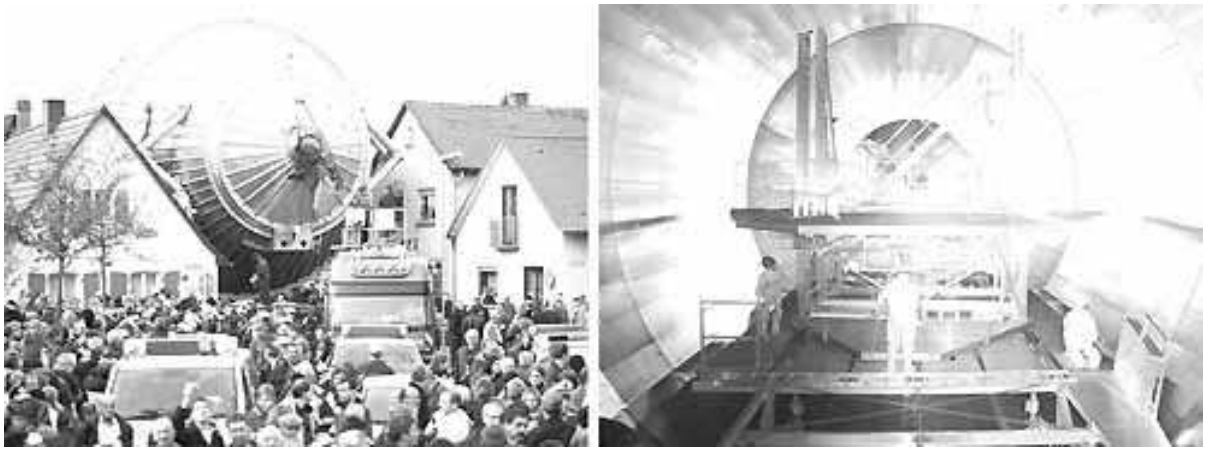}}
 \caption{Left: in November 2006 the main spectrometer vessel ($200\,t$) arrived
 at FZK after a $9000\,km$ long journey. Right: a custom made adjustable cleanroom scaffolding
 has been set up inside the spectrometer to install the wire electrodes on the inner wall.}
 \label{fig-mainspec}
\end{figure}

The central part of the experiment will be the tandem spectrometer
section. It consists of two electrostatic spectrometers of
MAC-E-Filter type: the {\em pre-spectrometer}, allowing only
electrons with the highest energies to pass into the {\em main
spectrometer}, where their kinetic energy will be analysed with a
resolution of $0.93\, eV$. The high energy resolution of the main
spectrometer requires large dimensions (see
fig.~\ref{fig-mainspec}). It has a diameter of $10\, m$, an
overall length of $23.4\, m$, a surface area of $690\,m^2$ and a
volume of $1240\,m^3$. The {\bf pre-spectrometer} with a diameter
of $1.7\, m$, a length of $3.4\, m$, a surface of $25\, m{^2}$ and
a volume of $8.5\, m{^3}$ served as a prototype for the design of
the main spectrometer. It demonstrated the scalability of the
vacuum layout and the reliability of the large $1.7\, m$ flange
design at all temperatures.

During standard operation very good UHV conditions of $p\,<\,
10^{-11}\,mbar$ have to be maintained in both spectrometer vessels
at room temperature in order to keep a low background rate. The
vacuum systems are based on a combination of cascaded
turbo-molecular pumps (TMP) and NEG-pumps made of double-coated
$30\, mm$ wide Zr-V-Fe getter strips\footnote{SAES
St707\textsuperscript{\textregistered} {\bf n}on-{\bf e}vaporable
{\bf g}etter strips}. Both vessels are made of electro-polished
stainless steel (316LN). Hydrogen outgassing from the walls is
expected to be the main source of gas, which limits the final
pressure in each vessel. The surface of each vacuum vessel can be
baked at temperatures up to $350^\circ C$ for reduction of
outgassing and activation of the NEG-pumps. The temperature of the
walls is controlled by heat-transfer oil pumped through a system
of stainless steel pipes, welded to the surface of each vessel.

The pre-spectrometer vacuum measurements showed in 2004, that a
hydrogen outgassing rate of $10^{-12}\,mbar \, l/s \,cm^2$ can be
reached with 316LN stainless steel, using standard manufacturing
procedures \cite{Upsala2005}. The cleaning process after
manufacturing included pickling, electro-polishing and rinsing
with de-ionized water. Based on this measured outgassing rate the
main spectrometer requires a total pumping speed of $10^6\, l/s$,
which will be provided by $3000\, m$ of NEG strips
\cite{Day07},\cite{Luo04}.

For the main spectrometer similar production methods and cleaning
procedures were used. During manufacturing a large workshop hall
was reserved for KATRIN. Before manufacturing started the floor
was coated and the whole hall was cleaned. The distance between
the manufacturer\footnote{MAN-DWE in Deggendorf/Germany)} and FZK
is about $400\,km$. Due to the large size of the main
spectrometer, it had to be shipped on a river barge down the river
Danube to the black sea, around Europe and up the river Rhine on a
$9000\,km$ detour (see fig.\ref{fig-mainspec}, left). After
delivery and installation of the heating and vacuum systems in
first commissioning tests in 2007 confirmed an outgassing rate of
$10^{-12}\,mbar \, l/s \,cm^2$, as expected from pre-spectrometer
measurements.

The electrostatic field will be generated by connecting the outer
wall of each spectrometer to high voltage ($-18.6\,kV$). The field
will be fine-tuned by an inner electrode system, made of very thin
stainless steel wires. Both pre-spectrometer and main spectrometer
have large, metal-sealed vacuum flanges (diameter: $1.7\,m$),
which allow to enter the vessel for installation of the inner
electrode system. The beam-tubes end in cone-shaped ground
electrodes, which are attached to the vacuum vessels via conical
ceramic insulators. The shape of the inner electrodes have been
calculated in extensive Monte Carlo simulations and tested with
the pre-spectrometer. A lot of effort went into the investigation
of penning traps in the $\vec{E} x \vec{B}$ fields, which can lead
to a strong increase of background rates. Currently the
pre-spectrometer is used to optimize the electrode design.

The electrode system of the main spectrometer consists of 248
individual wire electrode frames with a total of 23120 stainless
steel wires, with each wire electrically insulated against the
frame. These electrodes are presently produced at the university
M\"unster. At the same time preparations have been started to
install these electrodes inside the main spectrometer under
cleanroom conditions (see fig.\ref{fig-mainspec}, right). The
electrodes will be installed in 2009.

Another challenge is the high voltage system of the spectrometer.
The variable retarding voltage has to be known with ppm accuracy.
Since no known commercially available precision high voltage
divider meets this requirement, a unique voltage divider with an
appropriate stability has been developed at the university
M\"unster together with PTB Braunschweig. Together with this high
voltage system the electro-magnetic properties of the spectrometer
will be investigated in an intensive test programme, which also
includes the segmented silicon detector, currently being built at
the university of Washington in Seattle.

\section{Conclusions}

The KATRIN experiment has ambitious goals, both in particle
physics and in the technical realization of the experimental
set-up. Currently all major components are either under
construction or have already been delivered to FZK. First vacuum
measurements with the large main spectrometer vessel indicate an
outgassing behaviour, which agrees well with results from earlier
measurements with the much smaller pre-spectrometer. After
installing the final vacuum system, we expect to reach the
required pressure of $p\, <\, 10^{-11}\, mbar$ in the spectrometer
section. Other components of the experiment face different
challenges, like the DPS, where TMPs have to be operated close to
super-conducting solenoids or the WGTS with its stringent
requirements on stability of temperature and pressure.

The next step will be the installation of the inner electrode
system of the main spectrometer in 2009 an a test of its energy
resolution. After delivery and test of all components we expect
first tritium measurements in 2012.


\section*{Acknowledgment}
This work has been supported in part by the German BMBF
(05CK5VKA/5, 05CK5PMA/0, 05CK1REA/1) and by the DFG project SFB
Transregio 27.

\end{document}